# Nonspecific transcription factor-DNA binding influences nucleosome occupancy in yeast

Ariel Afek[1#], Itamar Sela[1#], Noa Musa-Lempel[2], and David B. Lukatsky[1*],
[1]*Department of Chemistry, Ben-Gurion University of the Negev, Beer-Sheva 84105 Israel*
[2]*Department of Computer Science, Ben-Gurion University of the Negev, Beer-Sheva 84105 Israel*

## Abstract
Quantitative understanding of the principles regulating nucleosome occupancy on a genome-wide level is a central issue in eukaryotic genomics. Here, we address this question using budding yeast, *Saccharomyces cerevisiae*, as a model organism. We perform a genome-wide computational analysis of nonspecific transcription factor (TF)-DNA binding free energy landscape, and compare this landscape with experimentally determined nucleosome binding preferences. We show that DNA regions with enhanced nonspecific TF-DNA binding are statistically significantly depleted of nucleosomes. We suggest therefore that the competition between TFs with histones for nonspecific binding to genomic sequences might be an important mechanism influencing nucleosome-binding preferences *in vivo*. We also predict that poly(dA:dT) and poly(dC:dG) tracts represent genomic elements with the strongest propensity for nonspecific TF-DNA binding, thus allowing TFs to outcompete nucleosomes at these elements. Our results suggest that nonspecific TF-DNA binding might provide a barrier for statistical positioning of nucleosomes throughout the yeast genome. We predict that the strength of this barrier increases with the concentration of DNA binding proteins in a cell. We discuss the connection of the proposed mechanism with the recently discovered pathway of active nucleosome reconstitution.

[*] Corresponding author: lukatsky@bgu.ac.il
[#] These authors contributed equally to the work

## Introduction

The introduction of high-throughput methods for determining nucleosome organization across entire genomes has provided a new perspective on understanding and modeling the regulation of eukaryotic gene expression [1-12]. These studies have shown, first of all, that promoters are often depleted of nucleosomes compared with coding regions [1-7, 13, 14]. Second, functionally related genes share nucleosome occupancy patterns in their promoters [4, 15]. Third, it appears that at least for a fraction of genome, promoter regions of highly transcribed genes are more depleted of nucleosomes compared with promoter regions of repressed genes [2, 4]. The question of how nucleosome occupancy and regulation of gene expression are related, appears to be the most complex and challenging issue since first, seminal studies of this relationship [16-19]. This is partly due to the fact that there are multiple additional factors, such as chromatin remodelers, and the competition with transcription factors (TFs), influencing gene expression [20, 21].

Intrinsic DNA sequence preferences of nucleosomes, has been a long-standing question for more than three decades [22-28]. Yet, a general answer to this question at the genome-wide level is still debated and a matter of active research [5-7, 10]. It appears that there are two dominant sequence features for nucleosome positioning. First, nucleosomes are depleted from sequences enriched in poly(dA:dT) both *in vivo* and *in vitro* [4, 6, 7, 13, 19]. This depletion is significantly stronger *in vivo* than *in vitro* [6, 7]. Second, nucleosomes are preferably positioned in sequences with AA/TT/AT and GG/CC/CG dinucleotides repeated with a period of about 10 nucleotides [5-7, 22, 23]. The second sequence feature is observed to be stronger *in vitro* than *in vivo* [7], and overall, this periodicity shows a statistically weak signal [7, 22, 23].

Comparison of genome-wide measurements of nucleosome occupancy *in vivo* and *in vitro* suggests that in a large fraction of the yeast genome *in vivo*, predominantly outside promoter regions enriched in poly(dA:dT) tracts, nucleosome occupancy is not intrinsically determined by the sequence, but it can be rather interpreted using a statistical positioning model [7, 10, 11]. This model assumes the existence of physical barriers at specific genomic locations, inducing nucleosome ordering in the vicinity of such barriers [29]. It was shown recently that the *in vivo* nucleosome occupancy can be reconstructed in a cell extract, *in vitro*, in the presence of adenosine triphosphate (ATP) [10]. This discovery suggests that barriers for statistical positioning of nucleosomes operate in an ATP-facilitated manner. This raises a key question: what mechanism provides such physical barriers for statistical nucleosome positioning? The latter question has become even more mysterious after it was shown that the transcription initiation complex is not an obvious barrier against which nucleosomes are organized [10]. It was shown that specifically bound TFs might provide a barrier for statistical nucleosome positioning only for a limited fraction of the yeast genome, thus leaving the latter question open [6, 10, 11, 30, 31].

Here, we suggest that nonspecific TF-DNA binding might provide such barriers genome-wide. We show that the free energy of nonspecific TF-DNA binding regulates the nucleosome occupancy genome-wide in yeast, *in vivo*. In particular, genomic regions depleted in nucleosomes possess a significantly lower free energy of nonspecific TF-DNA binding than genomic regions enriched in nucleosomes.

We have recently predicted that DNA sequence correlations statistically regulate nonspecific TF-DNA binding preferences [32]. Intuitively, sequence correlations mean statistically significant repeats of sequence patterns within genomic DNA. In particular, we have shown that enhanced homo-oligonucleotide sequence correlations, such as poly(dA:dT) or poly(dC:dG) tracts, where nucleotides of the same type are clustered together, generically reduce the nonspecific TF-DNA binding free energy, thus enhancing the nonspecific TF-DNA binding affinity. Sequence correlations where nucleotides of different types are alternating, lead to an opposite effect, reducing the nonspecific TF-DNA binding affinity [32]. In this paper we use this method in order to compute the nonspecific TF-DNA free energy landscape genome-wide, and compare this landscape with the high-resolution *in vivo* nucleosome occupancy data from Ref. [4], (see Supporting Material, **Figure S1**).

This article is organized as follows. First, we compute the nonspecific TF-DNA free energy landscape of the yeast genome, using a variant of the Berg-von Hippel model for TF-DNA binding, developed previously [32, 33]. We show that the average free energy of nonspecific TF-DNA binding exhibits a striking correlation with the average nucleosome occupancy profile, **Figure 1**. Second, we show that the origin of the predicted effect stems from the sequence correlation properties of the yeast genome, **Figure 2**. Third, we discuss the nonspecific TF-DNA free energy profiles for individual genes genome-wide, **Figure 3** and **Figure 4**. Forth, we present a minimal thermodynamic model describing the competition between TF and histone-octamer binding to DNA. We show in conclusion that the positioning of +1 and +2 nucleosomes statistically significantly correlates with the minimum of the nonspecific TF-DNA binding free energy genome-wide, **Figure 5**. We also analyze the relationship between the transcriptional plasticity and the presence of specific transcription factor binding sites (TFBSs), and the nonspecific TF-DNA binding free energy profiles, **Figure 6**. Finally, we discuss the connection of the proposed mechanism with the recently discovered pathway of the ATP-facilitated nucleosome reconstitution [10], and propose experiments that will allow direct testing of the predicted effects.

## Results

### Free energy of nonspecific TF-DNA binding

We begin by computing the free energy landscape of nonspecific TF-DNA binding at the genome-wide scale. For our analysis, we use genomic sequences from a high-resolution nucleosome occupancy atlas, obtained for budding yeast, *S. cerevisiae*, grown in the YPD medium [4], **Figure S1**. Our key working hypothesis that we advocate here is that such nonspecific TF-DNA binding regulates nucleosome occupancy in yeast.

In order to compute the free energy of nonspecific TF-DNA binding, we use a variant of the classical Berg-von Hippel model [32, 33]. In particular, the binding energy of TF with the length $M$, at a given DNA site $i$:

$$U(i) = -\sum_{j=i}^{M+i-1}\sum_{\alpha=1}^{4} K_\alpha \sigma_\alpha(j), \qquad \text{Eq. (1)}$$

where $\sigma_\alpha(j)$ is a four-component vector of the type $(\delta_{\alpha A}, \delta_{\alpha T}, \delta_{\alpha C}, \delta_{\alpha G})$, specifying the identity of the basepair at each DNA position $j$, with $\delta_{\alpha\beta} = 1$, if $\alpha = \beta$, and $\delta_{\alpha\beta} = 0$, if $\alpha \neq \beta$ (see Supporting Material, Materials and Methods). For example, if a given DNA site, $j$, is occupied by the T nucleotide, this vector takes the form: $(0,1,0,0)$; if the site $j$ is occupied by the G nucleotide, this vector is $(0,0,0,1)$. Within the framework of our model, each TF is fully described by four energy parameters, $K_A$, $K_T$, $K_C$, and $K_G$ [32].

In order to model nonspecific TF-DNA binding, we generate an ensemble of TFs, where for each TF, we draw the energies $K_A$, $K_T$, $K_C$, and $K_G$ from the Gaussian probability distributions, $P(K_\alpha)$, with zero mean and standard deviations (SD), $\sigma_\alpha = 2k_BT$, where $\alpha = \mathrm{A,T,C,G}$. The chosen magnitude of SD represents a typical strength of a hydrogen or electrostatic bond between TF and a nucleotide in contact with this TF. We note that the adopted model is highly simplified, as it assumes that the energy contributions of individual basepairs to the total TF-DNA binding energy are additive. In addition, it assumes that the energy of each contact is exclusively defined by the basepair type. Yet we suggest, based on our earlier analytical analysis [32], that the conclusions obtained below are quite general, and most likely, represent the rule rather than the exception.

We stress that the introduced notion of nonspecific TF-DNA binding means that the ensemble of model TFs is not designed in any way to bind preferentially any particular sequence motifs, but it rather represents a pool of random DNA binders. This allows us, after averaging out with respect to the ensemble of TFs, to introduce a purely DNA sequence-dependent signature of the DNA propensity for nonspecific binding to DNA-binding proteins.

For each model TF, sliding along a particular DNA sequence, we compute the free energy, $F = -k_B T \ln Z$, of TF-DNA binding within a sliding window of the width $L$, with the partition function:

$$Z = \sum_{i=1}^{L} \exp\left(-U(i) / k_B T\right), \quad \text{Eq. (2)}$$

where $k_B$ is the Boltzmann constant, and $T$ is the effective temperature. Intuitively, the effective temperature has two contributions. The first one is a conventional thermodynamic temperature due to thermal fluctuations of molecules. This is the only contribution existing *in vitro*. The second contribution stems *in vivo* from non-equilibrium, active fluctuations of the ATP-dependent chromatin modifying factors [10] (see below).

In order to remove the bias stemming from the global variability of nucleotide composition in different genomic locations, it is natural to analyze the free energy difference between the actual sequences and their randomized analogs, $\Delta F = F - F_\infty$, where $F_\infty$ is the free energy computed for randomized sequence of the same width, $L$, and averaged over 50 random realizations, for a given TF. Each random realization preserves the average nucleotide composition in a given sliding window. We used $L = 50$ bp for the width of the sliding window in the calculations. We verified that the results are very weakly sensitive to the variation of $L$ (data not shown).

For each sliding window, we perform the free energy calculations for 256 TFs, and we average the free energies of all TFs, $\langle \Delta F \rangle_{TF}$. We then assign this average free energy, $\langle \Delta F \rangle_{TF}$, to the midpoint within the sliding window. Repeating this procedure in different sliding windows, thus allows us to assign the average free energy of nonspecific TF-DNA binding to each DNA sequence position within the entire genome. Next, we compute the average of the free energy with respect to all sequences aligned by their transcription start sites (TSSs) for each cluster, $\langle \langle \Delta F \rangle_{TF} \rangle_{seq}$, where $\langle ... \rangle_{seq}$, denotes the second averaging, **Figure 1**.

The results of the average free energy calculations, **Figure 1**, show a striking correlation with the average nucleosome occupancy in all four clusters of yeast genes. The lower the average free energy of nonspecific TF-DNA binding at a given DNA position, the lower is the average nucleosome occupancy at this position, **Figure 1**. We conclude, therefore, that nonspecific TF-DNA interactions significantly influence the nucleosome occupancy genome-wide in yeast. We suggest that the mechanism for nucleosome depletion is the competition of histones with TFs for nonspecific binding to DNA. In the regions of the reduced free energy of nonspecific binding, TFs simply outcompete nucleosomes. We will discuss this competition quantitatively in more details below. Most importantly, we suggest that nonspecifically bound TFs provide a barrier for statistical nucleosome positioning [29, 34]. Yet, in addition to the concept introduced in seminal works of Kornberg et al. [29, 34], here we propose that the strength of this dynamical barrier varies across the genome, **Figure 7**. It depends on the magnitude of the free energy of nonspecific TF-DNA binding at a given genomic location, on the overall concentration of TFs and other DNA-binding proteins, and on the effective temperature that we discuss below. In particular, Kornberg et al. have shown that the presence of a barrier alone is sufficient to induce the nucleosome ordering downstream of the barrier [29, 34]. Therefore, the proposed hypothesis that nonspecifically bound TFs provide such a barrier, can explain the experimentally observed periodicity of the nucleosome occupancy in the regions downstream of the TSSs, despite the fact that the free energy of nonspecific TF-DNA binding does not show the periodicity in this region, **Figure 1**.

A notable feature of the computed free energy profiles for clusters 2 and 3 (promoters in these clusters show the strongest affinity for nonspecific binding), is that nucleosomes are depleted beyond the regions of the reduced free energy, **Figure 1**. This observation is consistent with the observation of Field et al. [35] that nucleosomes are depleted over the boundaries of poly(dA:dT) tracts. The statistical nucleosome positioning mechanism by nonspecifically bound TFs, that we propose here, might provide an explanation for this observation, since the free energy reaches its minimum in regions enriched in poly(dA:dT) and poly(dC:dG) tracts. Namely, a diffusion cloud of TFs extends beyond the regions of enhanced nonspecific TF-DNA binding, such as poly(dA:dT) and poly(dC:dG) tracts. Intuitively, we expect that the lower the free energy of nonspecific TF-DNA binding characterizing a particular DNA stretch, the wider is the nucleosome-depleted region around this stretch. We discuss this effect quantitatively in more details below.

**DNA sequence correlations control propensity for nonspecific TF-DNA binding**

In this section we provide a connection between the correlation properties of DNA sequences and the free energy of nonspecific TF-DNA binding, genome-wide in yeast. We have recently shown analytically that enhanced homo-oligonucleotide sequence correlations, such as poly(dA:dT) and poly(dC:dG) tracts, statistically enhance the nonspecific TF-DNA binding affinity [32]. We predicted that generically, nonspecific TF-DNA binding is controlled by the strength and symmetry of DNA sequence correlations, and our conclusions are qualitatively robust with respect to the details of the microscopic model for TF-DNA binding [32]. We now show that the free energy landscape computed above is consistent with the genome-wide correlation landscape of the yeast genome.

Statistical correlation properties of DNA sequences can be described by the normalized correlation function [32] (Materials and Methods),

$$C_{\alpha\alpha}(x) = s_{\alpha\alpha}(x) / \left\langle s^{r}_{\alpha\alpha}(x) \right\rangle , \qquad \text{Eq. (3)}$$

where $s_{\alpha\alpha}(x) = \left\langle \sigma_{\alpha}(i)\sigma_{\alpha}(i+x) \right\rangle$, and $\left\langle s^{r}_{\alpha\alpha}(x) \right\rangle$ is obtained analogously, using the set of randomly permuted sequences averaged with respect to different random realizations. This definition, Eq. (3), removes the average nucleotide compositional bias and thus describes the correlation properties of different nucleotide types on the same footing, despite the compositional differences between nucleotides in different genomic regions. To characterize the sequence correlations further, we introduce the cumulative correlation function:

$$\mathrm{H}_{\alpha\alpha} = \sum_{x=1}^{x_{\max}} C_{\alpha\alpha}(x) , \qquad \text{Eq. (4)}$$

where the summation is performed with respect to the first $x_{\max}$ values of $C_{\alpha\alpha}(x)$ (Materials and Methods). The larger the magnitude of $\mathrm{H}_{\alpha\alpha}$, the stronger are the homo-oligonucleotide correlations.

We computed $\mathrm{H}_{\alpha\alpha}$ for four clusters of yeast genes; **Figure 2** shows the results for $\mathrm{H}_{\mathrm{TT}}$. The results for $\mathrm{H}_{\mathrm{AA}}$ are similar to $\mathrm{H}_{\mathrm{TT}}$; and $\mathrm{H}_{\mathrm{CC}}$ and $\mathrm{H}_{\mathrm{GG}}$ show also similar, yet weaker correlation profiles (data not shown). In agreement with our generic, analytical predictions [32], the cumulative correlation profiles exhibit an anti-correlation with the free energy profiles, **Figure 1**, and with the nucleosome occupancy. The stronger the homo-oligonucleotide correlations in a specific genomic location, the lower is the free energy of nonspecific TF-DNA binding, and thus the lower is the nucleosome occupancy in this location. A key prediction that follows from our analysis is that, as far as the nonspecific TF-DNA binding affinity is concerned, poly(dC:dG) tracts should act similarly to poly(dA:dT) tracts. In a seminal work of Iyer and Struhl [19] it was shown that for the HIS3 promoter region, poly(dC:dG) induces an effect similar to poly(dA:dT), facilitating the gene expression. Here we suggest that this effect is quite general, and it stems from enhanced nonspecific TF-DNA binding induced by enhanced homo-oligonucleotide sequence correlations, represented, in particular, by extended poly(dA:dT) and poly(dC:dG) tracts.

**Single-gene free energy profiles**

At the single-gene level, there is a high degree of variability in the correlation between the free energy of nonspecific TF-DNA binding and the nucleosome occupancy, **Figure 3**. Single-gene free energy profiles show a significantly higher degree of variability than the corresponding nucleosome occupancy profiles, **Figure 3**. The overall linear correlation coefficients between the nucleosome occupancy and $\langle \Delta F \rangle_{TS}$ at the single-gene level, computed within the range $(-250,250)$ around the TSS, are low but statistically significant: $R \simeq 0.17$, $R \simeq 0.18$, $R \simeq 0.14$, and $R \simeq 0.20$ for clusters 0, 1, 2, and 3, respectively, with the $p$-value, $p = 0$, for all clusters. Changing the range around the TSS where the correlation coefficient is computed, does not lead to a significant variation in $R$ for either cluster (data not shown). We conclude therefore, that the striking correspondence between the average nucleosome occupancy and $\langle \langle \Delta F \rangle_{TS} \rangle_{seq}$, **Figure 1**, does not persist at the single-gene level, where $\langle \Delta F \rangle_{TS}$ varies significantly from gene-to-gene, even within the same cluster. Examples of the correspondence between the free energy of nonspecific TF-DNA binding and the nucleosome occupancy for eight genes from different clusters are shown in **Figure 4**. The HIS3 gene represents a classical case, studied in a seminal work of Iyer and Struhl [19]. It is remarkable that in the upstream promoter region of HIS3, $\langle \Delta F \rangle_{TS}$ exhibits a very good correlation with the nucleosome occupancy profile, including a local minimum at TSS, where the TATA box elements are located, and a distant, local minimum within the region $(-350,-300)$. On the contrary, in the promoter region of highly regulated GAL1 gene, the -1 nucleosome is positioned at -50, and the nucleosome occupancy is not well correlated with $\langle \Delta F \rangle_{TS}$ in this region.

To further characterize the effect of nonspecific TF-DNA binding on nucleosome occupancy, we ordered all genes according to their minimal value of the free energy, $\Delta F_{\min} = \min[\langle \Delta F \rangle_{TS}]$, within the interval $(-150,0)$. For each gene, we also computed the distance, $D$, between $\Delta F_{\min}$ and the maximal value of the nucleosome occupancy, $\max(NO)$, within the interval $(0,150)$, where the +1 nucleosome is positioned (see the last figure in **Figure 4** for the graphical definition of $D$). We then divided the genes into bins according to the value of $\Delta F_{\min}$, and computed the average, $\langle \Delta F_{\min} \rangle$ and $\langle D \rangle$ in each bin. The correlation plot $\langle D \rangle$ versus $\langle \Delta F_{\min} \rangle$ is shown in **Figure 5.A**. The lower the free energy minimum, $\langle \Delta F_{\min} \rangle$, the larger is $\langle D \rangle$. In order words, the stronger the nonspecific TF-DNA binding in the promoter region, the further away from $\Delta F_{\min}$ the +1 nucleosome is positioned. We also computed the standard deviation, $\sigma_D$, of $D$ in each bin, **Figure 5.B**. We observe that the lower $\langle \Delta F_{\min} \rangle$, the smaller is $\sigma_D$. Both of these observations can be understood intuitively in the following way. We suggest that the +1 nucleosome is statistically positioned by the barrier formed by the cloud of TFs nonspecifically bound to the promoter region. The lower the free energy, $\langle \Delta F_{\min} \rangle$, the more pronounced is this effect (i.e. the higher is the potential barrier), and thus the further away the +1 nucleosome is positioned, **Figure 7**. Intuitively, one expects that the degree of positional variability, $\sigma_D$, increases with the weakening of the barrier strength. This is indeed consistent with our result, **Figure 5.B**.

We also repeated this analysis for the +2 nucleosome, located downstream of the +1 nucleosome, **Figure 5.C** and **D**. We again observed that the distance between $\Delta F_{\min}$ and the maximum of the nucleosome occupancy for the +2 nucleosome (see **Figure 4**, CDC11) are correlated with $\langle \Delta F_{\min} \rangle$, **Figure 5.C** and **D**, yet this correlation is weaker than the corresponding correlation for the +1 nucleosome, **Figure 5.A** and **B**. Remarkably, we did not observe any statistically significant correlation for the distance between $\Delta F_{\min}$ and the maximal occupancy for the -1 nucleosome, located upstream of the TSS (data not shown). This is not surprising, since the average maximal nucleosome occupancy of the -1 nucleosome is about three times smaller compared to the +1 or +2 nucleosomes, and in addition, it is much weaker positioned (high fuzziness) [4, 7, 10]. Taken together, these observations suggest that nonspecific TF-DNA binding provides a statistical barrier in the upstream vicinity of the TSS, and significantly influences the nucleosome positioning downstream of the TSS.

Barrier-induced ordering (often termed 'wall-induced crystallization') is a well-known concept in statistical mechanics [36, 37]. It is well established that the presence of a potential barrier dramatically reduces the free energy barrier for crystal nucleation [36, 37], which provides a direct analogy for nucleosome positioning induced by a barrier. Yet, a quantitative understanding of the effect of varying barrier strength on nucleosome positioning is an open question for a future investigation.

### Minimal thermodynamic model of competition between TFs and nucleosomes

We now present a minimal thermodynamic model that describes the competition between histone complexes and TFs for binding to genomic DNA. The reactions that describe the process:

$$
\begin{aligned}
&[N]+[D] \underset{k_{-N}}{\overset{k_N}{\rightleftarrows}} [N\bullet D] \\
&[T]+[D] \underset{k_{-T}}{\overset{k_T}{\rightleftarrows}} [T\bullet D]
\end{aligned}
\qquad \text{Eq. (5)}
$$

where $[N]$, $[T]$, and $[D]$, denote the concentrations of free histone octamers (*i.e.* free nucleosomes), TFs, and DNA, respectively; and $[N\bullet D]$ and $[T\bullet D]$ denote the concentrations of nucleosomes and TFs, respectively, bound to DNA. Here, we assume for simplicity that the binding occurs only between histone octamers and DNA, and there is no binding between histone monomers, dimers, and tetramers with DNA. We also assume that all TFs are indistinguishable with respect to their nonspecific binding preferences towards DNA, and therefore the overall concentration of TFs is much larger than the concentration of nucleosomes, $[T] \gg [N]$ (we estimated that, on average, the overall concentration of TFs in yeast is at least 20 times larger than the concentration of histone octamers). The solution of the kinetic equations in the steady state gives the equilibrium concentrations of $[N\bullet D]$ and $[T\bullet D]$:

$$
\begin{aligned}
X &= \frac{\chi_T}{\chi_N + \chi_T + \chi_N \chi_T} \\
Y &= \frac{\chi_N}{\chi_N + \chi_T + \chi_N \chi_T}
\end{aligned}
\qquad \text{Eq. (6)}
$$

where $X = [N \bullet D]/[D_{tot}]$, $Y = [T \bullet D]/[D_{tot}]$, $\chi_N = K_N / [N]$, and $\chi_T = K_T / [T]$; $[D_{tot}]$ is the total cellular concentration of genomic DNA accessible to TFs. The equilibrium dissociation constants, $K_N$ and $K_T$, are expressed as functions of the corresponding binding free energies:

$$K_N = C_0^N \exp(\Delta F_N / k_B T)$$
$$K_T = C_0^T \exp(\Delta F_T / k_B T) \quad \text{Eq. (7)}$$

where $\Delta F_N$ and $\Delta F_T$ are the average binding free energies of nonspecific binding of histone octamers and TFs to DNA, respectively; $C_0^N$ and $C_0^T$ are the inverse molecular volumes of nucleosome and TF, respectively.

Taking into account that $[T] \gg [N]$, the fraction of bound nucleosomes is:

$$X \simeq \frac{C_0^T [N]}{C_0^N [T]} \exp\left([\Delta F_T - \Delta F_N] / k_B T\right). \quad \text{Eq. (8)}$$

Therefore, the nucleosome occupancy, $NO \sim \log_2 X$, measured in experiments [4], takes the form:

$$NO \sim (\Delta F_T - \Delta F_N) / k_B T . \quad \text{Eq. (9)}$$

The predicted effect of nonspecific binding is exclusively entropic, $F = \langle U \rangle - TS \simeq -TS$, since we have shown recently [32], $\langle U \rangle \simeq 0$. Assuming that the mobility of a large nucleosome along DNA is negligible compared to a TF, the effect of nonspecific nucleosome-DNA binding is negligible, $|\Delta F_N| \ll |\Delta F_T|$. While the latter assumption is definitely valid *in vitro* (and *in vivo* in the absence of ATP) [38], yet its validity is uncertain in the presence of ATP, *in vivo* [39]. Adopting this assumption as being valid, the key conclusion here is that the nucleosome occupancy predicted from the simplest thermodynamic model follows the profile of nonspecific TF-DNA binding free energy, $NO \sim \Delta F_T / k_B T$, **Figure 1**.

We note that the presented model is highly simplified and it takes into account only the effect of nonspecific, competitive interactions of TFs and histone octamers with DNA. This model does not take into account steric effects, stemming from a finite nucleosome size, and the effect of enhanced, structural repulsion of nucleosomes by poly(dA:dT) tracts. In addition, we assumed that TFs compete for binding to DNA only with histone octamers, while in reality histone monomers, dimers, and presumably higher-order histone complexes also compete with histone octamers [40]. Taking into account this competition should make the predicted effect even stronger. Finally, taking into account the nucleosome-induced cooperativity between TFs [41], may further improve the accuracy of the predictions.

In summary, we presented the simplest possible model that takes into account the competition of TFs with histone octamers for DNA binding. This model predicts that the average nucleosome occupancy profile should simply follow the free energy profile for nonspecific TF-DNA binding, **Figure 1**. The key, open issue is to explain the experimentally observed periodicity of the nucleosome occupancy in the regions downstream of the TSSs, despite the fact that the free energy of nonspecific TF-DNA binding does not show the periodicity in this region, **Figure 1**. We suggest that the discrepancy between the actual, *in vivo* nucleosome occupancy profile and our simplified model prediction can be reconciled, if one takes into account the effect of a steric barrier induced by nonspecifically bound TFs, providing a statistical nucleosome positioning. In their seminal works, Kornberg et al. have shown that the presence of such a barrier alone is sufficient to induce the nucleosome ordering downstream of the barrier [29, 34]. Here we suggest that nucleosomes are periodically organized against the dynamic barrier provided by nonspecifically bound TFs. Further elucidation of the latter issue is the subject of our future work.

# Discussion

### Summary and discussion of key findings

Here we predicted the molecular nature of a barrier for statistical positioning of nucleosomes in yeast. We suggest that nonspecific TF-DNA binding provides such barrier *in vivo*, on a genome-wide basis. We predict, quite generally, that TFs (and other DNA-binding proteins) compete with histones for nonspecific DNA binding in different genomic locations, thus influencing nucleosome occupancy genome-wide. Most importantly, we predict quantitatively the landscape of such nonspecific TF-DNA binding free energy for the yeast genome. Strikingly, we find a strong correlation between the average free energy and the average nucleosome occupancy profiles genome-wide, **Figure 1**.

The molecular origin of the predicted nonspecific TF-DNA binding stems from DNA sequence correlations. This effect is generic; it depends exclusively on the symmetry and the length-scale of DNA sequence correlations, and we predict that it is qualitatively robust with respect to microscopic details of the model [32]. We predict, in particular, that poly(dA:dT) and poly(dC:dG) tracts represent genomic elements with the strongest propensity for nonspecific TF-DNA binding. The action of such nonspecific binding potential allows TFs to outcompete histones in the promoter regions of the majority of genes in the yeast genome. We predict that TFs, nonspecifically bound to promoter regions in the upstream vicinity of TSS, may provide a barrier for statistical positioning of +1, +2, and further nucleosomes, downstream of the TSS, genome-wide, **Figure 7**. The proposed effect is statistically strong due to the fact that the predicted binding potential acts nonspecifically on all TFs (and other DNA-binding proteins). In particular, for a typical TF making a contact with ten nucleotide basepairs, the predicted free energy of nonspecific binding in the upstream vicinity of the TSS, is within the range between $-1\,k_B T$ and $-5\,k_B T$, on average, **Figure 4**, that corresponds to a few kcal, which is a strong effect.

The predicted effect does not exclude the current model that poly(dA:dT) tracts intrinsically disfavor nucleosome formation; and also, it is not in contradiction with an experimentally proven fact that specifically bound TFs provide a barrier for statistical nucleosome positioning at some genomic locations [7, 10, 30, 31, 34]. Here we suggest that in addition to the known mechanisms, nonspecific TF-DNA binding is an important factor influencing nucleosome occupancy *in vivo*. The fact that there is a high variability of the nonspecific free energy landscapes for individual genes belonging to the same cluster, **Figure 3** and **Figure 4**, suggests that specific and nonspecific TF-DNA binding might be tightly linked. In order to explore this question, we computed the profiles of nonspecific TF-DNA binding free energy for the groups of genes with different numbers of specific TF binding sites in their promoter regions, **Figure 6.C**. We also computed the free energy profiles for the groups of genes characterized by the high and low transcriptional plasticity, respectively, **Figure 6.A**. We observe that both of these functional criteria are translated in notable differences in the free energy profiles, **Figure 6.A** and **C**, suggesting that nonspecific TF-DNA binding does have functional consequences in yeast. Further investigation of such interplay between specific and nonspecific binding is the subject for our future investigation.

One of the key findings of this investigation is the observed correlation between the minimum of the free energy of nonspecific TF-DNA binding and the positioning of the +1 and +2 nucleosomes, **Figure 5**. This observation provides a support for our working hypothesis that nonspecifically bound TFs constitute a barrier for statistical nucleosome positioning. The strength of this dynamic barrier varies across the genome; it depends on the correlation properties of DNA at a given genomic location, and on the overall concentration of TFs and other DNA-binding proteins. We suggest that the presence of such barriers induces the experimentally observed nucleosome ordering genome-wide, in genomic locations downstream of the TSSs.

The effect that we predict here is entropy driven, and it is thus enhanced at higher temperatures; i.e. the free energy of nonspecific TF-DNA binding is reduced at higher temperatures [32]. We note that strictly speaking, the temperature, $T$ , entering the definition of the free energy, $F = -k_B T \ln Z(T)$, Eq. (2), is not a conventional thermodynamic temperature, when the *in vivo* system is considered. It is rather an effective, 'active' temperature, which is a measure of active fluctuations produced by the presence of (adenosine triphosphate) ATP-dependent chromatin modifying factors. The notion of an effective, 'active' temperature is a well-established concept in non-equilibrium statistical mechanics [42]. The degree of such active fluctuations varies across the genome. It depends on the sequence at specific genomic locations, on the expression level and localization of chromatin modifying factors and TFs, and on the concentration of ATP. Therefore, unlike the conventional thermodynamic temperature that acts uniformly on the entire cell, the active temperature is a complex, dynamic function, which is a subject to tight regulation in a living cell.

**ATP-dependent, active nucleosome reconstitution**

The entropic mechanism for nonspecific TF-DNA binding proposed in this paper, combined with the notion of the 'active' temperature, can explain, at least qualitatively, a recently discovered pathway for ATP-dependent, active nucleosome reconstitution [10]. In particular, it was observed that the *in vivo* pattern of nucleosome occupancy in yeast, can be reproduced *in vitro*, using a whole-cell extract, exceptionally in the presence of ATP [10].

Our results can qualitatively explain the key observation of Ref. [10] if we adopt a model that ATP-dependent chromatin modifying factors simply increase the effective temperature of the system, by producing strong, 'active' fluctuations. As a result of such fluctuations, the predicted barrier induced by nonspecifically bound TFs becomes more pronounced, **Figure 7**, thus enhancing the statistical nucleosome positioning. Our key hypothesis here is that the predicted barrier for statistical nucleosome positioning is a dynamic, entropically driven barrier, which depends on the effective temperature of the system at specific genomic locations. The presence of such barriers is sufficient to induce the experimentally observed nucleosome ordering, genome-wide *in vivo*, in genomic locations downstream of the TSSs [29, 34]. A further support of our hypothesis comes from a recent experimental study [12], which demonstrated that the deletion of the key chromatin remodeling factors, ISW1, ISW2, and CHD1, severely disrupts the nucleosome ordering genome wide in yeast. The observed disruption of the nucleosome ordering is the strongest in the triple-deletion mutant [12]. We suggest that the observed effect of these mutations, at least partially, can be interpreted as simply reducing the active temperature of the system, thus reducing the strength of the dynamic barrier. A more quantitative investigation of this key issue is also the subject for our future study.

We note finally that our results are also qualitatively consistent with the striking observation that nucleosome turnover rates are the fastest in promoter regions of yeast, and not in the coding regions [39]. We suggest that a dominant mechanism for the nucleosome turnover in promoter regions might be a competition with nonspecifically bound TFs. Since the predicted nonspecific TF-DNA binding is enhanced in promoter regions compared to coding regions, we might expect that this entropy dominated effect will lead to 'hotter' nucleosomes in promoters, as it was indeed observed in [39].

**Proposed experiments**

We conclude by proposing experiments that will allow direct testing of the predicted effect. A key experiment would measure the modulation of nucleosome occupancy, combined with measurements of TF-DNA binding genome-wide, upon insertion of poly(dA:dT) or poly(dC:dG) tracts of variable length into promoters of genes. The design of this experiment is conceptually similar to the design of Iyer and Struhl presented in their seminal paper for the case of HIS3 promoter [19]. ChIP-chip or ChIP-seq methods allow genome-wide measurements of both nucleosome occupancy and TF-DNA binding for hundreds of TFs [43]. We expect that at longer poly(dA:dT) or poly(dC:dG) tracts, nonspecific TF-DNA binding will be enhanced, thus leading to a more pronounced nucleosome depletion. Our theoretical results suggest that a barrier for statistical nucleosome positioning, induced by nonspecifically bound TFs, will be stronger upon increasing the overall concentration of TFs and other DNA binding proteins in a cell. The expression level of TFs can be systematically modulated genome-wide using the existing overexpression libraries [44]. We predict that overexpression of TFs will generically lead to a more pronounced depletion of nucleosomes in promoter regions enriched in poly(dA:dT) and poly(dC:dG), and to a stronger barrier for statistical nucleosome positioning, leading to a higher degree of nucleosome ordering induced by the presence of this barrier.

In yeast, a direct experimental test of the predicted effect is further complicated by the fact that the majority of its promoters are A/T-rich, which leads to an enhanced structural repulsion of nucleosomes from such regions, compared to G/C-rich regions that intrinsically favor nucleosome formation [45]. In order to decouple DNA structural effects from the predicted entropic effect of nonspecific protein-DNA binding on nucleosome binding preferences, we suggest to systematically vary the correlation properties of G/C-rich regions in yeast, keeping their average G/C content fixed. We expect that increasing the fraction of even short poly(dC:dG) tracts (of up to 4 bp long), should significantly decrease the nucleosome occupancy of such poly(dC:dG)-rich regions, compared to uncorrelated G/C-rich regions with exactly the same average G/C content.

## Acknowledgements

We thank O. Rando, E.I. Shakhnovich, K. Struhl, I. Tirosh, and M. Ziv-Ukelson for helpful discussions. We thank I. Tirosh for kindly providing us the data on transcriptional plasticity. D.B.L. acknowledges the financial support from the Israel Science Foundation (ISF) grant 1014/09. A.A. is a recipient of the Lewiner graduate fellowship.

## Supporting Material

Materials and Methods, references [46,47], and one figure:
http://download.cell.com/biophysj/mmcs/journals/0006-3495/PIIS000634951101201X.mmc1.pdf

## References

1. **Bernstein, B. E., C. L. Liu, …, S. L Schreiber. 2004. Global nucleosome occupancy in yeast. *Genome Biol.* 5(9):R62.**
2. **Lee, C. K., Y. Shibata, …, J. D. Lieb. 2004. Evidence for nucleosome depletion at active regulatory regions genome-wide. *Nat Genet.* 36(8):900-905.**

3. Yuan, G. C., Y. J. Liu, ..., O. J. Rando. 2005. Genome-scale identification of nucleosome positions in S. cerevisiae. *Science*. 309(5734):626-630.
4. Lee, W., D. Tillo, ..., , C. Nislow. 2007. A high-resolution atlas of nucleosome occupancy in yeast. *Nat Genet*. 39(10):1235-1244.
5. Segal, E., Y. Fondufe-Mittendorf, ..., J. Widom. 2006. A genomic code for nucleosome positioning. *Nature*. 442(7104):772-778.
6. Kaplan, N., I. K. Moore, ..., E. Segal. 2009. The DNA-encoded nucleosome organization of a eukaryotic genome. *Nature*. 458(7236):362-366.
7. Zhang, Y., Z. Moqtaderi, ..., K. Struhl. 2009. Intrinsic histone-DNA interactions are not the major determinant of nucleosome positions in vivo. *Nat Struct Mol Biol*. 16(8):847-852.
8. Ioshikhes, I.P., I .Albert, ..., B. F. Pugh. 2006. Nucleosome positions predicted through comparative genomics. *Nat Genet*. 38(10):1210-1215.
9. Tirosh, I., N. Sigal, N. Barkai. 2010. Divergence of nucleosome positioning between two closely related yeast species: genetic basis and functional consequences. *Mol Syst Biol*. 6:365.
10. Zhang, Z., C. J. Wippo, ..., B. F. Pugh. 2011. A packing mechanism for nucleosome organization reconstituted across a eukaryotic genome. *Science.* 332(6032):977-980.
11. Mavrich, T. N., I. P. Ioshikhes, B. F. Pugh. 2008. A barrier nucleosome model for statistical positioning of nucleosomes throughout the yeast genome. *Genome Res.* 18(7):1073-1083.
12. Gkikopoulos, T., P. Schofield, ..., T. Owen-Hughes. 2011. A role for Snf2-related nucleosome-spacing enzymes in genome-wide nucleosome organization. *Science.* 333(6050):1758-1760.
13. Sekinger. E.A., Z. Moqtaderi, K. Struhl. 2005. Intrinsic histone-DNA interactions and low nucleosome density are important for preferential accessibility of promoter regions in yeast. *Mol Cell.* 18(6):735-748.
14. Fan, X.C., Z. Moqtaderi, ..., K. Struhl. 2010.Nucleosome depletion at yeast terminators is not intrinsic and can occur by a transcriptional mechanism linked to 3 '-end formation. *P Natl Acad Sci USA.* 107(42):17945-17950.
15. Tirosh, I., N. Barkai. 2008. Two strategies for gene regulation by promoter nucleosomes. *Genome Res.* 18(7):1084-1091.
16. Fedor, M. J., R. D. Kornberg. 1989. Upstream activation sequence-dependent alteration of chromatin structure and transcription activation of the yeast GAL1-GAL10 genes. *Mol Cell Biol.* 9(4):1721-1732.
17. Svaren, J., W. Horz. 1997. Transcription factors vs nucleosomes: regulation of the PHO5 promoter in yeast. *Trends Biochem Sci.* 22(3):93-97.
18. Cavalli, G., F. Thoma. 1993. Chromatin transitions during activation and repression of galactose-regulated genes in yeast. *EMBO J.* 12(12):4603-4613.
19. Iyer, V., K. Struhl. 1995. Poly(dA:dT), a ubiquitous promoter element that stimulates transcription via its intrinsic DNA structure. *EMBO J.* 14(11):2570-2579.
20. Segal, E., J. Widom. 2009.From DNA sequence to transcriptional behaviour: a quantitative approach. *Nat Rev Genet.* 10(7):443-456.
21. Rando, O. J., H. Y. Chang . 2009. Genome-wide views of chromatin structure. *Annu Rev Biochem.* 78:245-271.
22. Trifonov, E.N., J. L. Sussman. 1980. The pitch of chromatin DNA is reflected in its nucleotide sequence. *Proc Natl Acad Sci U S A.* 77(7):3816-3820.

23. Ioshikhes, I., A. Bolshoy, ..., E. N. Trifonov. 1996. Nucleosome DNA sequence pattern revealed by multiple alignment of experimentally mapped sequences. *J Mol Biol.* 262(2):129-139.
24. Satchwell, S. C., H. R. Drew, A. A. Travers. 1986. Sequence periodicities in chicken nucleosome core DNA. *J Mol Biol.* 191(4):659-675.
25. Widom, J. 2001. Role of DNA sequence in nucleosome stability and dynamics. *Q Rev Biophys.* 34(3):269-324.
26. Segal, E., J. Widom. 2009. Poly(dA:dT) tracts: major determinants of nucleosome organization. *Curr Opin Struct Biol.* 19(1):65-71.
27. Miele, V., C. Vaillant, ..., T. Grange . 2008. DNA physical properties determine nucleosome occupancy from yeast to fly. *Nucleic Acids Res.* 36(11):3746-3756.
28. Scipioni A., S. Morosetti , P. De Santis. 2009. A Statistical Thermodynamic Approach for Predicting the Sequence-Dependent Nucleosome Positioning Along Genomes. *Biopolymers* . 91(12):1143-1153.
29. Kornberg, R. D., L. Stryer.1988. Statistical distributions of nucleosomes: nonrandom locations by a stochastic mechanism. *Nucleic Acids Res.* 16(14A):6677-6690.
30. Bai, L., A. Ondracka, F. R. Cross. 2011.Multiple sequence-specific factors generate the nucleosome-depleted region on CLN2 promoter. *Mol Cell.* 42(4):465-476.
31 Ganapathi, M., M. J. Palumbo, ..., R. H. Morse. 2010.Extensive role of the general regulatory factors, Abf1 and Rap1, in determining genome-wide chromatin structure in budding yeast. *Nucleic Acids Res.* 39(6):2032-2044.
32. Sela I, D. B. Lukatsky. 2011.DNA sequence correlations shape nonspecific transcription factor-DNA binding affinity. *Biophys J.* 101(1):160-166.
33. Berg, O.G., P. H. von Hippel. 1987. Selection of DNA binding sites by regulatory proteins. Statistical-mechanical theory and application to operators and promoters. *J Mol Biol.* 193(4):723-750.
34. Fedor, M. J., N. F. Lue, R. D. Kornberg. 1988. Statistical positioning of nucleosomes by specific protein-binding to an upstream activating sequence in yeast. *J Mol Biol.* 204(1):109-127.
35. Field, Y., N. Kaplan, ..., E. Segal. 2008. Distinct modes of regulation by chromatin encoded through nucleosome positioning signals. *PLoS Comput Biol.* 4(11):e1000216.
36. Auer, S., D. Frenkel. 2003. Line tension controls wall-induced crystal nucleation in hard-sphere colloids. *Phys Rev Lett.* 91(1):015703.
37. Cacciuto, A., S. Auer, D. Frenkel. 2004. Onset of heterogeneous crystal nucleation in colloidal suspensions. *Nature.* 428(6981):404-406.
38. Pazin M.J., P. Bhargava, ..., J.T. Kadonaga. 1997. Nucleosome mobility and the maintenance of nucleosome positioning. *Science.* 276(5313):809-812.
39. Dion M.F., T. Kaplan, ..., O.J. Rando. 2007. Dynamics of replication-independent histone turnover in budding yeast. *Science.* 315(5817):1405-1408.
40. Andrews, A. J., X. Chen, ..., K. Luger. 2010. The histone chaperone Nap1 promotes nucleosome assembly by eliminating nonnucleosomal histone DNA interactions. *Mol Cell.* 37(6):834-842.
41. Mirny, L.A. 2010.Nucleosome-mediated cooperativity between transcription factors. *Proc Natl Acad Sci U S A.* 107(52):22534-22539.
42. Cugliandolo, L. F., J. Kurchan, L. Peliti. 1997.Energy flow, partial equilibration, and effective temperatures in systems with slow dynamics. *Phys Rev E.* 55(4):3898-3914.

43. **Venters, B.J., S. Wachi, …,B. F. Pugh. 2011. A comprehensive genomic binding map of gene and chromatin regulatory proteins in Saccharomyces. *Mol Cell.* 41(4):480-492.**
44. **Sopko, R., D. Huang, …,B. Andrews . 2006. Mapping pathways and phenotypes by systematic gene overexpression. *Mol Cell.* 21(3):319-330.**

45. **Tillo D, Hughes TR. 2009. G+C content dominates intrinsic nucleosome occupancy. *BMC Bioinformatics*. 10:442.**
46. **Ihmels, J., G. Friedlander, …, N. Barkai. 2002. Revealing modular organization in the yeast transcriptional network. *Nat Genet.* 31(4):370-377.**
47. **MacIsaac, K. D., T. Wang, …, E. Fraenkel. 2006. An improved map of conserved regulatory sites for Saccharomyces cerevisiae. *BMC Bioinformatics.* 7:113.**

## Figure Legends

**Figure 1.** Computed average free energy of nonspecific TF-DNA binding (blue), $\langle \Delta f \rangle = \langle \langle \Delta F \rangle_{TF} \rangle_{seq} / M$, normalized per bp. The contact energies, $K_\alpha$, were drawn from the Gaussian distribution, $P(K_\alpha)$, with zero mean, $\langle K_\alpha \rangle = 0$, and standard deviation, $\sigma_\alpha = 2k_BT$. The averaging, $\langle ... \rangle_{TF}$, for each sliding window is performed over 256 TFs, and over all sequences, $\langle ... \rangle_{seq}$, in a given cluster (Materials and Methods). In order to compute error bars, we divided each cluster into five sub-clusters, and computed $\langle \Delta f \rangle$ for each sub-cluster. The error bars are defined as one standard deviation of $\langle \Delta f \rangle$ between sub-clusters. The nucleosome occupancy data, *NO*, from Ref. [4], measured with the 4 bp resolution, are shown for comparison (red).

**Figure 2.** Computed cumulative correlation functions, $H_{TT}$, for each cluster (Materials and Methods). The error bars are computed similarly to **Figure 1**.

**Figure 3.** Heat maps of the nonspecific TF-DNA binding free energies (normalized per bp), $\Delta f = \langle \Delta F \rangle_{TF} / M$, computed for single-genes, for each cluster. Heat maps of nucleosome occupancy are also shown for comparison.

**Figure 4.** Single-gene examples. The nonspecific TF-DNA binding free energy (normalized per bp) profiles, $\Delta f = \langle \Delta F \rangle_{TF} / M$ (blue) computed for eight individual genes. The corresponding nucleosome occupancy profiles are shown for comparison (red). The last example (CDC11) shows the position of the minimum of $\Delta f$, and the maxima of the nucleosome occupancy of +1 and +2 nucleosomes, respectively.

**Figure 5. A.** Correlation plot of the average distance, $\langle D \rangle$, between the minimum, $\Delta f_{\min}$, of $\Delta f(x)$ in the range (-150,0) and the maximum of the nucleosome occupancy in the range (0,150) for +1 nucleosome (see CDC11 gene example in **Figure 4**). All 5,014 genes from all clusters are grouped into 23 bins (218 genes in each bin). $D$ and $\Delta f_{\min}$ is computed for each individual gene in each bin, and then the averages, $\langle D \rangle$ and $\langle \Delta f_{\min} \rangle$ are computed in each bin. **B.** The standard deviation, $\sigma_D$, of $D$ for individual genes, is computed in each bin, as a function of $\langle \Delta f_{\min} \rangle$. **C.** The average distance, $\langle D \rangle$, between the minimum, $\Delta f_{\min}$, of $\Delta f(x)$ in the range (-150,0) and the maximum of the nucleosome occupancy in the range (150,300), corresponding to +2 nucleosome. **D.** The standard deviation, $\sigma_D$, of $D$ for individual genes for +2 nucleosome is computed in each bin, as a function of $\langle \Delta f_{\min} \rangle$. All free energies are normalized per bp, as above.

**Figure 6. A.** The average free energy, $\langle \Delta f \rangle = \langle \langle \Delta F \rangle_{TF} \rangle_{seq} / M$, for nonspecific TF-DNA binding (normalized per bp), computed for the top 10% of genes with the lowest (green) and highest (violet) transcriptional plasticity, respectively. There are 488 genes in each of these two groups. **B.** The average nucleosome occupancy profiles from Ref. [4], for the corresponding two groups of genes, used in computing (**A**). In order to compute error bars, we divided each group into four sub-groups, and computed $\langle \Delta f \rangle$ for each sub-group. The error bars are defined as one standard deviation of $\langle \Delta f \rangle$ between the sub-groups. **C.** The average free energy, $\langle \Delta f \rangle$, computed for the groups of genes with a different number of known TFBSs in promoter regions. The number of genes in each group: 2648 (0), 1302 (1), 538 (2), 227 (3), and 169 (4+). **D.** The average nucleosome occupancy profiles from Ref. [4], for the corresponding groups of genes, used in computing **(C)**.

**Figure 7.** Schematic representation of the predicted mechanism for statistical nucleosome positioning by nonspecifically bound TFs. Nucleosomes are represented as large, blue ovals; transcription factor as small, red ovals. The free energy landscape of nonspecific TF-DNA binding (blue curves) leads to a weak (top) and strong (bottom) attraction of TFs, respectively. In the former case, the barrier for nucleosome positioning is weak (nucleosomes are not well localized); while in the latter case, the barrier is strong (well-localized, periodically ordered nucleosomes).

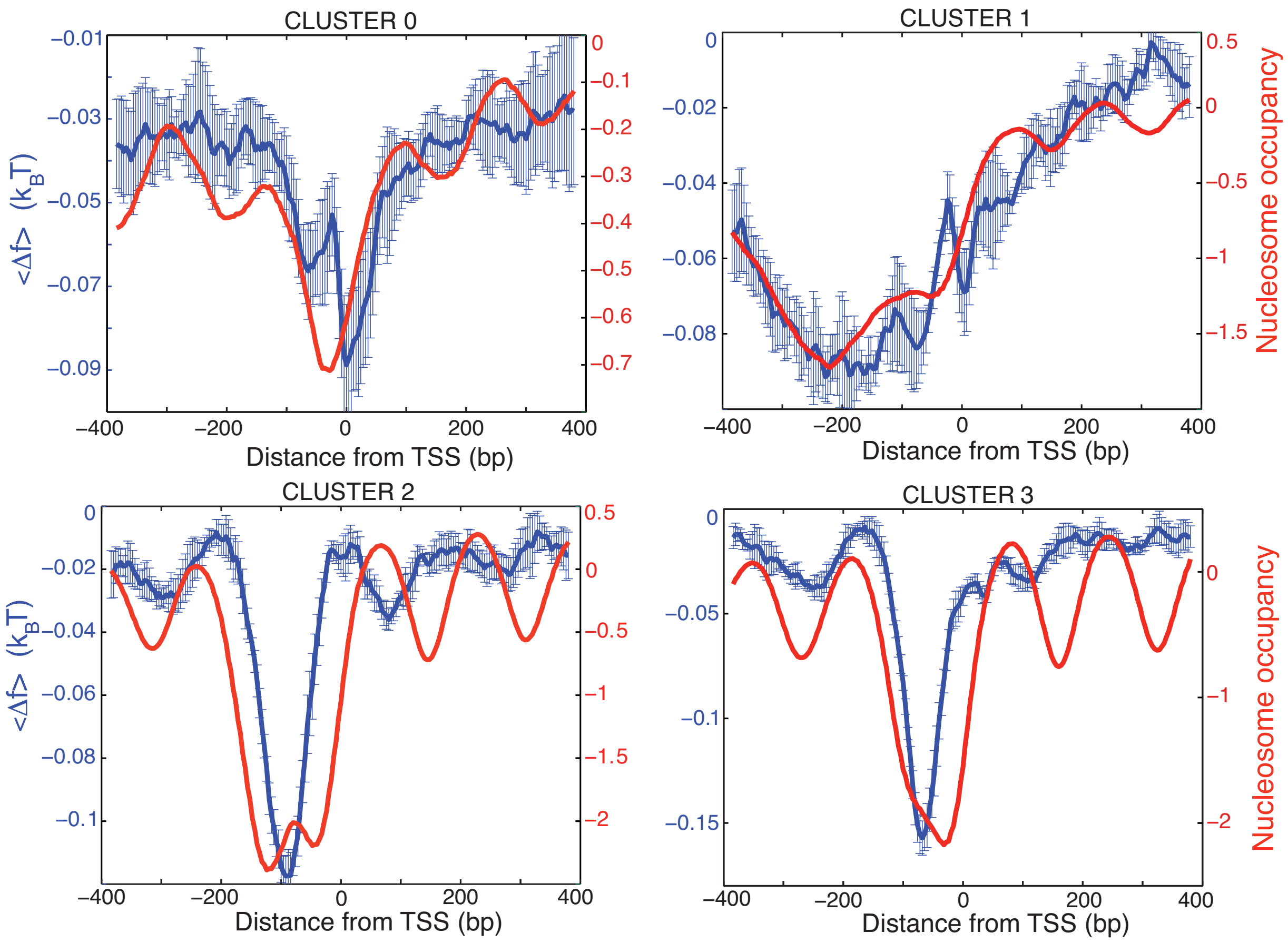

**Figure 1.**

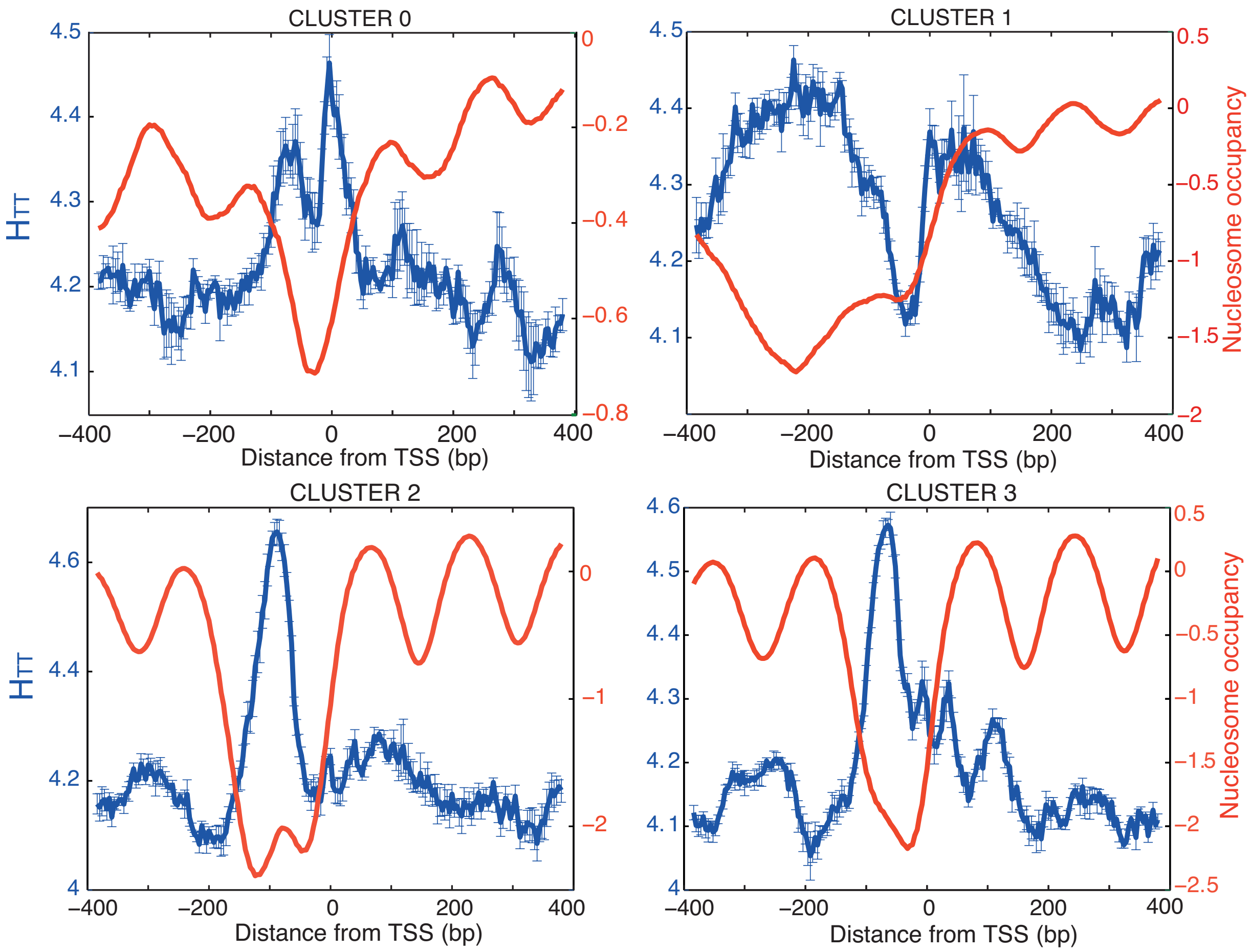

**Figure 2.**

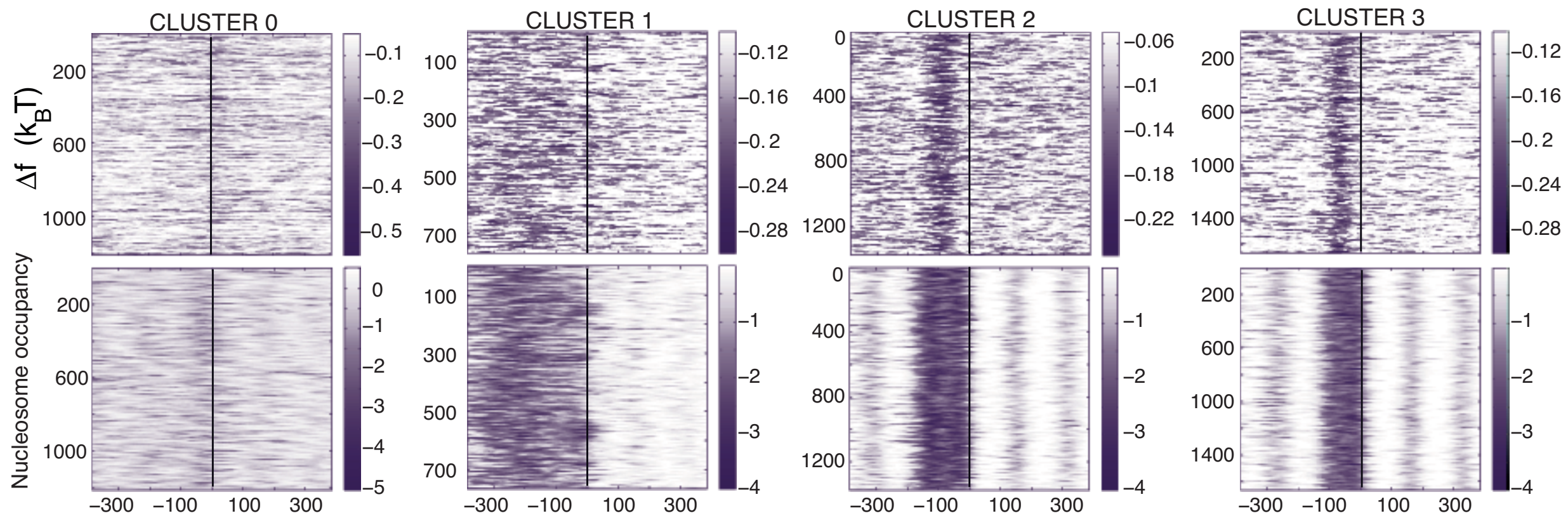
**Figure 3.**

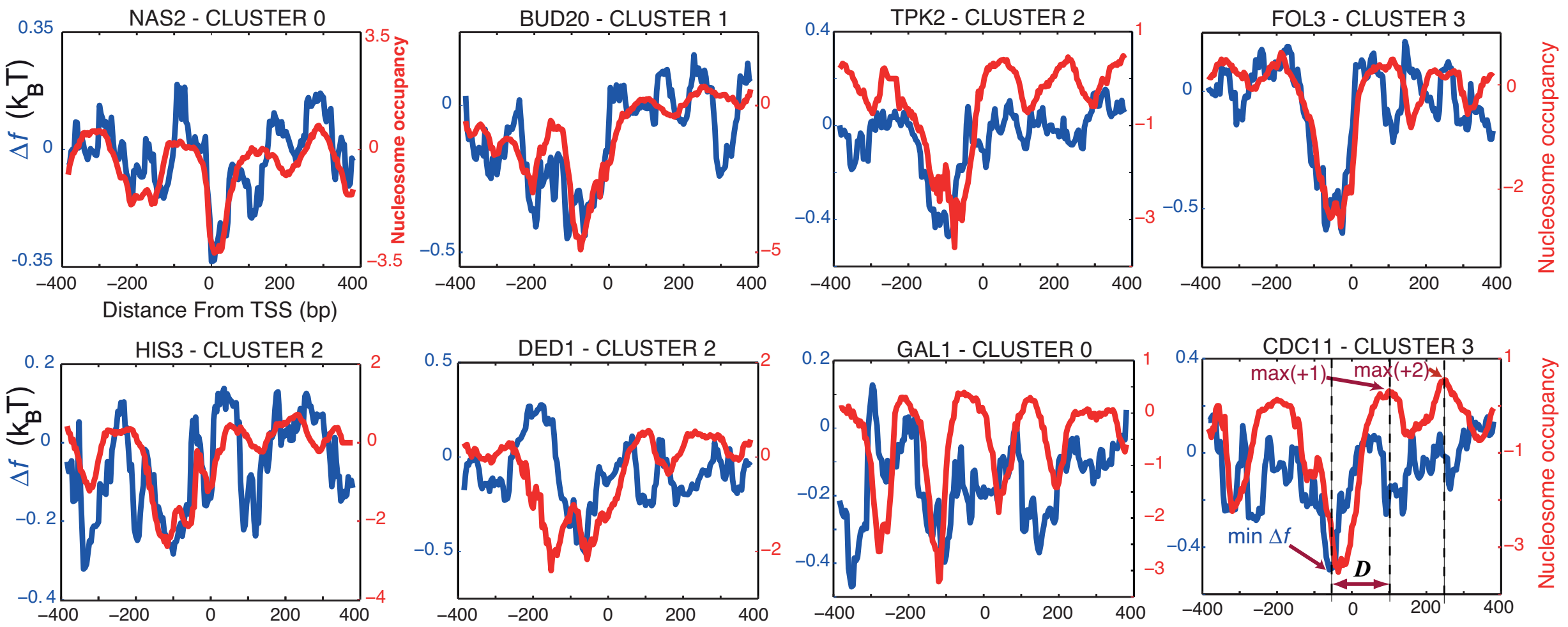
NAS2 - CLUSTER 0
BUD20 - CLUSTER 1
TPK2 - CLUSTER 2
FOL3 - CLUSTER 3
HIS3 - CLUSTER 2
DED1 - CLUSTER 2
GAL1 - CLUSTER 0
CDC11 - CLUSTER 3
Δf (kBT)
Nucleosome occupancy
Distance From TSS (bp)
max(+1)
max(+2)
min Δf
D

**Figure 4.**

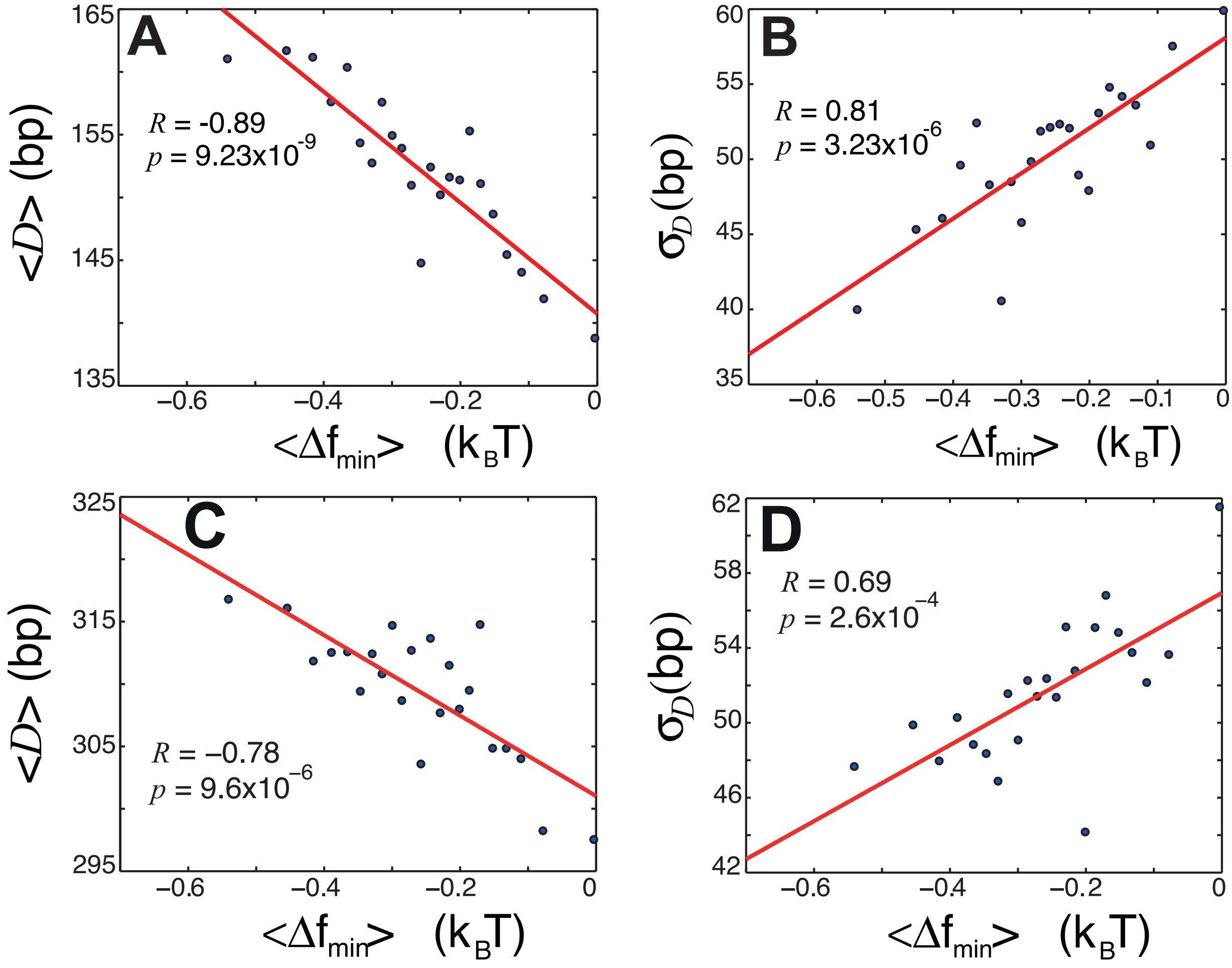

**Figure 5.**

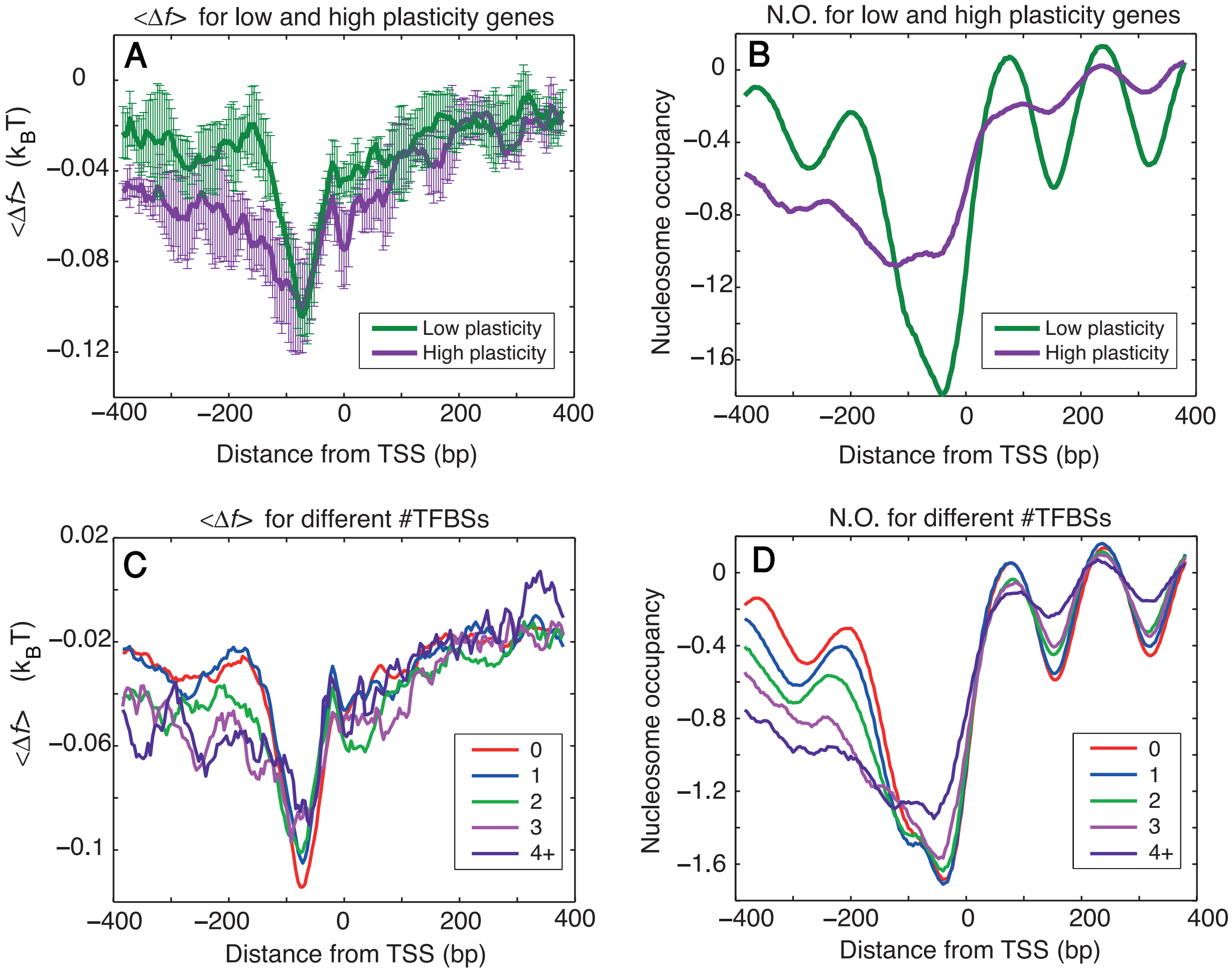

**Figure 6.**

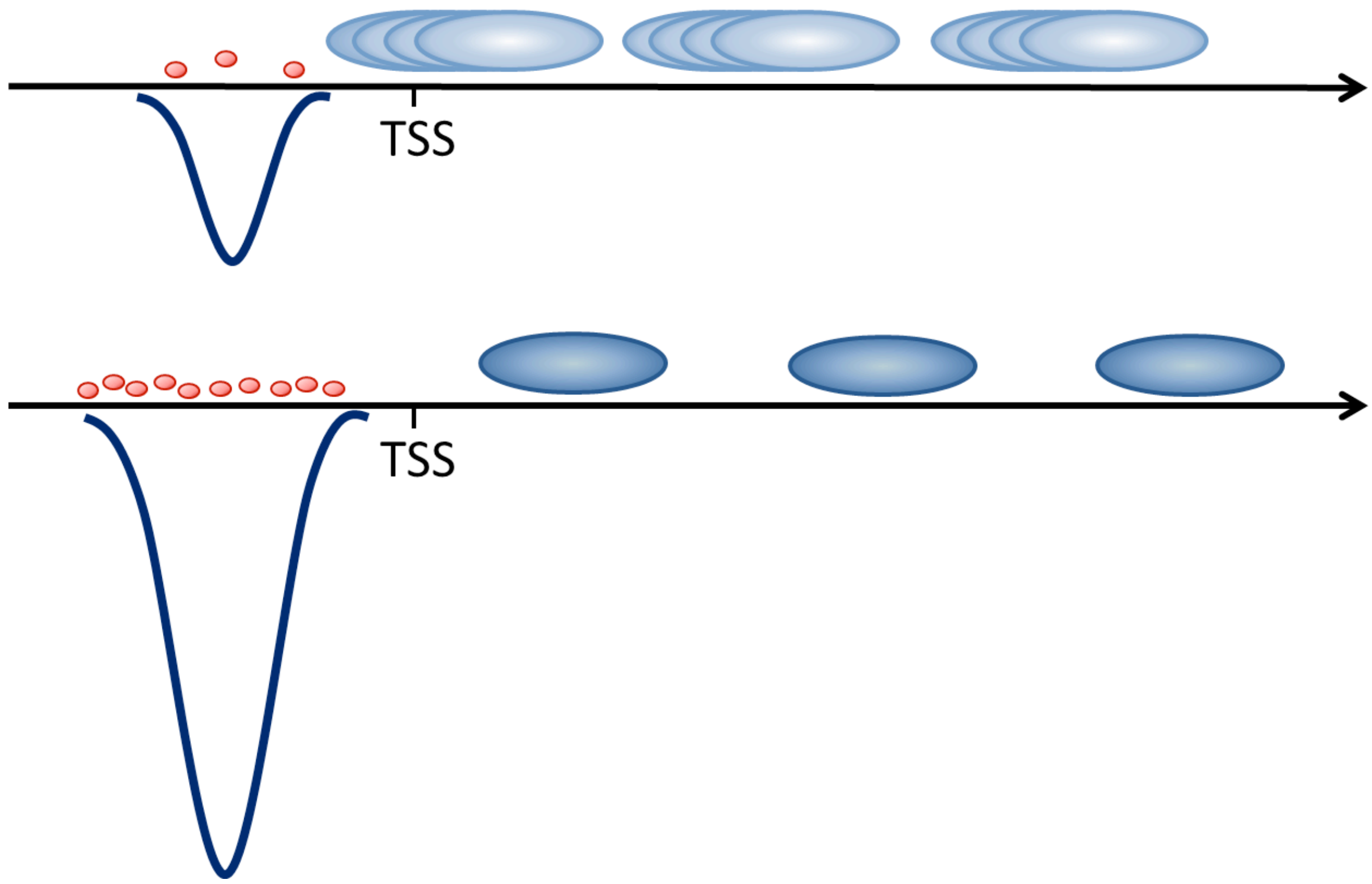

**Figure 7.**

# Supporting Material - Nonspecific transcription factor-DNA binding influences nucleosome occupancy in yeast

## Materials and Methods

### Free energies of nonspecific TF-DNA binding genome-wide

We used the following procedure in order to assign the magnitude of the nonspecific TF-DNA binding free energy at each genomic location, $x_0$, within the interval $(-384;380)$ around the TSS for each of 5,014 yeast genes, distributed into four clusters. First, for a given gene, we position the middle of the sliding window of the width $L = 50$ bp at $x_0$. Second, for each TF (out of 256 randomly generated TFs) we compute the partition function, Eq. (2), and the free energy, $F = -k_B T \ln Z$, in a given sliding window. In order to get rid of the compositional bias in different genomic locations, we always compute the difference, $\Delta F = F - F_\infty$, where $F_\infty$ is computed for randomized sequence of the same width, $L$, in the same sliding window, and averaged over 50 random realizations, for a given TF. The randomization procedure permutes the positions of nucleotides, preserving the average nucleotide composition in the sliding window. Third, we repeat the latter procedure for each of 256 TFs, and average the free energies of all TFs at the genomic location, $x_0$, of this gene, $\langle \Delta F \rangle_{TF}$. Fourth, we move the sliding window to the next genomic location, $x_0 + \Delta x$, and repeat the entire procedure in order to obtain $\langle \Delta F \rangle_{TF}$ in this new location. In our analysis we computed $\langle \Delta F \rangle_{TF}$ in steps of 4 bp, $\Delta x = 4$. The contact energies of TF-DNA interactions are generated for each of 256 TFs, as described in the main text. We used $\sigma_\alpha = 2k_B T$ for the standard deviation of $P(K_\alpha)$, Eq. (1). The obtained free energy landscape is very weakly sensitive to the change of the sliding window width, $L$ (data not shown). We used, quite arbitrary, $M = 8$, for the length of each TF. Our conclusions are qualitatively robust for a wide range of $M$ (data not shown).

### Cumulative correlation functions genome-wide

We used the following procedure in order to compute the correlation function, $C_{\alpha\alpha}(x)$, Eq. (3), for each of four nucleotide types, $\alpha$. For each cluster of genes, aligned with respect to the TSS, we compute $C_{\alpha\alpha}(x)$ in a sliding window of width $L = 50$ bp, with the middle of the sliding window being positioned at $x_0$. In order to compute $\langle s^r_{\alpha\alpha}(x) \rangle$, we randomly permute the sequences in a given window, preserving the average nucleotide composition of each sequence. We used 50 random realizations in order to compute the average, $\langle s^r_{\alpha\alpha}(x) \rangle$. Next, we compute the cumulative correlation function, $\mathrm{H}_{\alpha\alpha}$, Eq. (4), at this location $x_0$, where the summation is performed up to a cutoff, $x_{\max} = 4$ bp. A different choice of the cutoff leads to a rescaling of the base level of $\mathrm{H}_{\alpha\alpha}$, without affecting the trend (data not shown). We then repeat the entire procedure for the next sliding window, $x_0 + \Delta x$, thus assigning the value of $\mathrm{H}_{\alpha\alpha}$ at each genomic location around the TSS, for a given cluster of genes. The obtained cumulative correlation functions, $\mathrm{H}_{\alpha\alpha}$, are very weakly sensitive to the change of the sliding window width, $L$ (data not shown).

### Transcriptional plasticity and TFBSs

In order to compute **Figure 6.A** and **B**, we used the classification of transcriptional plasticity from Refs. [1, 2]. In order to compute **Figure 6.C** and **D**, we used the TFBSs data from Ref. [3].

## Supporting References

1. Tirosh, I., N. Barkai. 2008. Two strategies for gene regulation by promoter nucleosomes. *Genome Res.* 18(7):1084-1091.
2. Ihmels J, Friedlander G, Bergmann S, Sarig O, Ziv Y, Barkai N. 2002. Revealing modular organization in the yeast transcriptional network. *Nat Genet*. 31(4):370-377.
3. MacIsaac KD, Wang T, Gordon DB, Gifford DK, Stormo GD, Fraenkel E. 2006. An improved map of conserved regulatory sites for Saccharomyces cerevisiae. *BMC Bioinformatics*. 7:113.

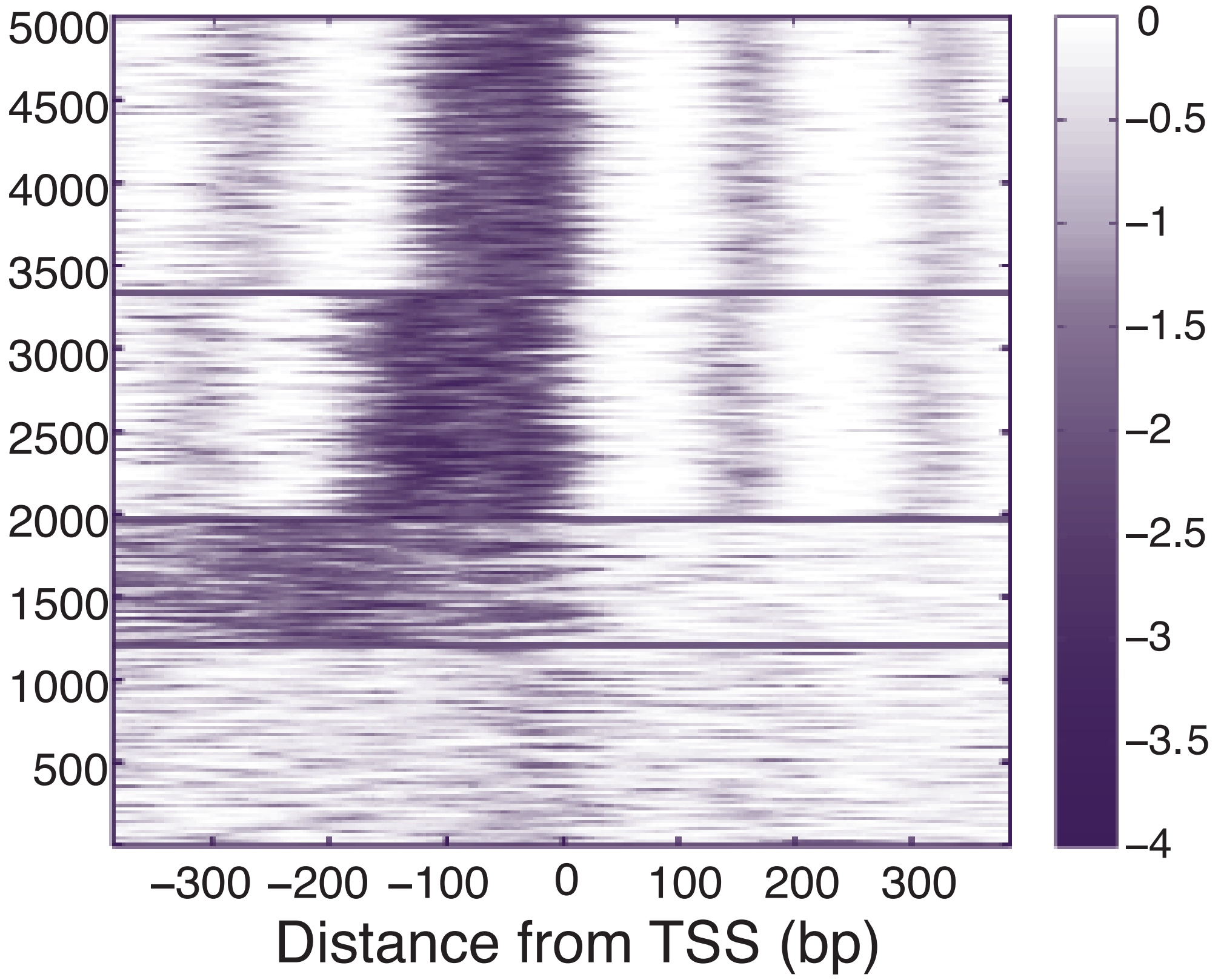

**Figure S1.** Graphical representation of the yeast nucleosome occupancy data, which we used for our analysis (see Figure 3 of Lee, W. *et al.* 2007. *Nat Genet*. 39(10):1235-1244). Average nucleosome occupancy, covering ~800 bp surrounding the TSS for 5,014 transcripts. The transcripts are clustered into four clusters according to the GO classification. Each cluster contains (from bottom to top): 1,211 stress response genes (cluster 0); 766 translation genes (cluster 1); 1,374 ribosome biogenesis and assembly genes (cluster 2); 1,663 organelle organization and biogenesis genes (cluster 3).